# Machine learning methods for turbulence modeling in subsonic flows over airfoils

Zhang Weiwei, Zhu Linyang, Liu Yilang, Kou Jiaqing
School of aeronautics, Northwestern Polytechnical University, Xi'an 710072, China

Reynolds-Averaged Navier-Stokes (RANS) method for high Reynolds number (Re) turbulent flows will still play a vital role in the following several decadein aerospace engineering. Although RANS models are widely used, empiricism and large discrepancies between models reduce the reliability of simulating complex flows. Therefore, in recent years, data-driven turbulence model has aroused widespread concern in fluid mechanics. Based on the experimental/numerical simulation results, this approach aims to modify or construct the turbulence model for specific purposes by machine learning techniques. The effectiveness of this method has been preliminarily verified for low Reynolds number turbulent flows based on direct numerical simulation (DNS) data. In this paper, we take the results calculated by Spallart-Allmaras (SA) model as training data and explore the feasibility of data-driven methods for high Reynolds number turbulent flows. Different from low Reynolds number turbulent flows, the data from high Reynolds number flows shows an apparent scaling effect, thus leading to difficulties in the data-driven modeling. In order to improve the fitting accuracy, we divided the flow field into near-wall region, wake region, and far-field region, and built individual model for every region. In this paper, we adopted the radial basis function neural network (RBFNN) and some auxiliary optimization algorithms to reconstruct a mapping function between mean variables and the eddy viscosity. Since this model reflects the relationship between local flow characteristics and turbulent eddy viscosity, it is independent on the airfoil shape and flow condition. The training data in this paper is generated from only three subsonic flow calculations of NACA0012 airfoil. By coupling the proposed approach with Navier-Stokes equations, we calculated various flow cases as well as two different airfoils (NACA0014 and RAE2822 airfoil) and showed the eddy viscosity contours, velocity profiles along the normal direction of wall and skin friction coefficient distributions, etc. Compared with the SA model, the results show a reasonable accuracy and better efficiency, which indicates the positive prospect of data-driven methods in turbulence modeling.

## 1. Introduction

In 1883, Reynolds discovered the turbulent state in pipe flow, marking the start of turbulent flow research. Based on the contribution of many predecessors, people now get more and more profound physical insight into turbulent flows. However, the essence of turbulence, how to control and use turbulent flows more efficiently are still suspending. At present, numerical simulation and experiments are the main sources to obtain turbulence results for engineering problems.But generally, high Reynolds number experiments in aeronautics are not only expensive and time-consuming, but also pretty hard to achieve elaborate measurement especially in boundary layer. The numerical methodscan be further classified as direct numerical simulation (DNS), large eddy simulation (LES) and Reynolds averaged Navier-Stokes simulation (RANS)according to different grid resolutions.

In computational fluid dynamics (CFD), with the improvement of computational capability, high-fidelity methods, like DNS and LES, have been increasingly used in turbulence computations, and have got some achievement in some practical contexts. However, DNS is still impractical in aeronautical industries due to the extremely high grid resolution, which is exponentially proportional to the Reynolds number. Besides, the aircraft simulation across the full flight envelope by LES needs more than high-performance computing (HPC) advances and improvements in algorithmic technology, which might not be realized until 2030 [1]. *Hence, RANS models will remain a critical approach in engineer practice during the foreseeable future* [2].

Traditional turbulent models mainly include algebraic models and transport models. Algebraic models (like Baldwin-Lomax (BL) model [3]) are simple but have lower accuracy, while transport models (e.g., SA model, $k-\varepsilon$ model, $k-\omega$ model, etc.) are more accurate but require the solution of partial differential equations (PDEs) [4]. All of these models are extensively adopted in general engineering applications because of their high efficiency and easy implementation. However, the universality of these models is limited since some prescribed parameters should be determined a priori, which are derived from some specific experiments and DNS results. *Moreover, for some complex separated flows, large discrepancies between turbulence models may occur, which force the users to choose the appropriate turbulence model according to their experience and the specific problem of interest.* Although Reynolds stress transport models (RSTM) can obtain higher accuracy, its complicated transport equations and poor convergence lead to less popularity in practical application.

The shortcomings of RANS models mentioned above are difficult to overcome from traditional studies. But recently, some new technologies based on data mining and machine learning have shown their potential in solving these problems. In turbulence, Milano and Koumoutsakos [5] approximated the high order terms by neural networks. Hoceva et al. [6] modeled some turbulent variables in airfoil wake region by radial basis function neural networks (RBFNN). In fact, the research work that formally adopts data-driven method to improve or replace RANS model is mainly carried out in the past five years [7]. Duraisamy et al. [8-11] modeled the source terms in SA turbulence model by neural networks and embedded it into CFD solver. The study above is mainly focused on the modeling of turbulence related variables. The purpose of the other research is to reduce the uncertainties between high fidelity data and results of RANS models. For example, Duraisamy and Singh et al. [12-14] combined the inversion model and machine learning to infer and reconstruct better functional forms in turbulence and transitions modeling; Xiao et al. [15-19] proposed the concept of "physics-informed machine learning (PIML)" to emphasize the importance of including the physical domain knowledge into machine learning. Different from these studies, other works are based purely on high fidelity data rather than classic RANS models, which further extended the application of machine learning to turbulence modeling and verified the positive prospect of data-driven methods. For example, Ling and Templeton et al. [20-23] embedded the invariance property into deep neural networks firstly and demonstrated the advantage over the architecture without this property. Different from the two kinds of study, Ling's work was based purely on high fidelity data rather than classic RANS models, which further developed the application of machine learning to turbulence modeling and verified the positive prospect of data-driven methods. Similarly, Gamabara and Hattori [24] adopted artificial neural network (ANN) to model the subgrid-scale stress in LES. A detailed description about turbulence modeling with data-driven techniques is reviewed in [25].

From the past three decades, the development of traditional RANS models has reached a plateau and is hard to achieve essential change and improvement. It is perhaps time for the turbulence modeling community to adopt challenge datasets [26]. *Different from those work to improve baseline RANS models or target to Reynolds stress modeling of low/middle Reynolds number flows, this paper underlines the practicality of reconstructing function form of eddy viscosity in high Reynolds number airfoil flows.* It should also be noted that, since the current approach does not need to solve PDEs, lower computational cost than transport models is expected. The rest of this paper is organized as follows. Section 2 introduces the methods of model constructing, including the sample selection, model framework and optimization of the model parameters. Section 3 characterizes the training and predicting datasets and the numerical results. At last, conclusions and future outlook are addressed in Section4.

# 2. Method

## 2.1. Modeling process

*The proposed approach to reconstruct eddy viscosity function can be divided into two parts: the learning machine and the surrogate machine*. The learning machine mainly includes sample selection, model framework and parameter optimization. In surrogate machine part, the proposed model is inserted into CFD solver, then the eddy viscosity according to mean flow variables are calculated and passed to CFD solver, see figure 1.

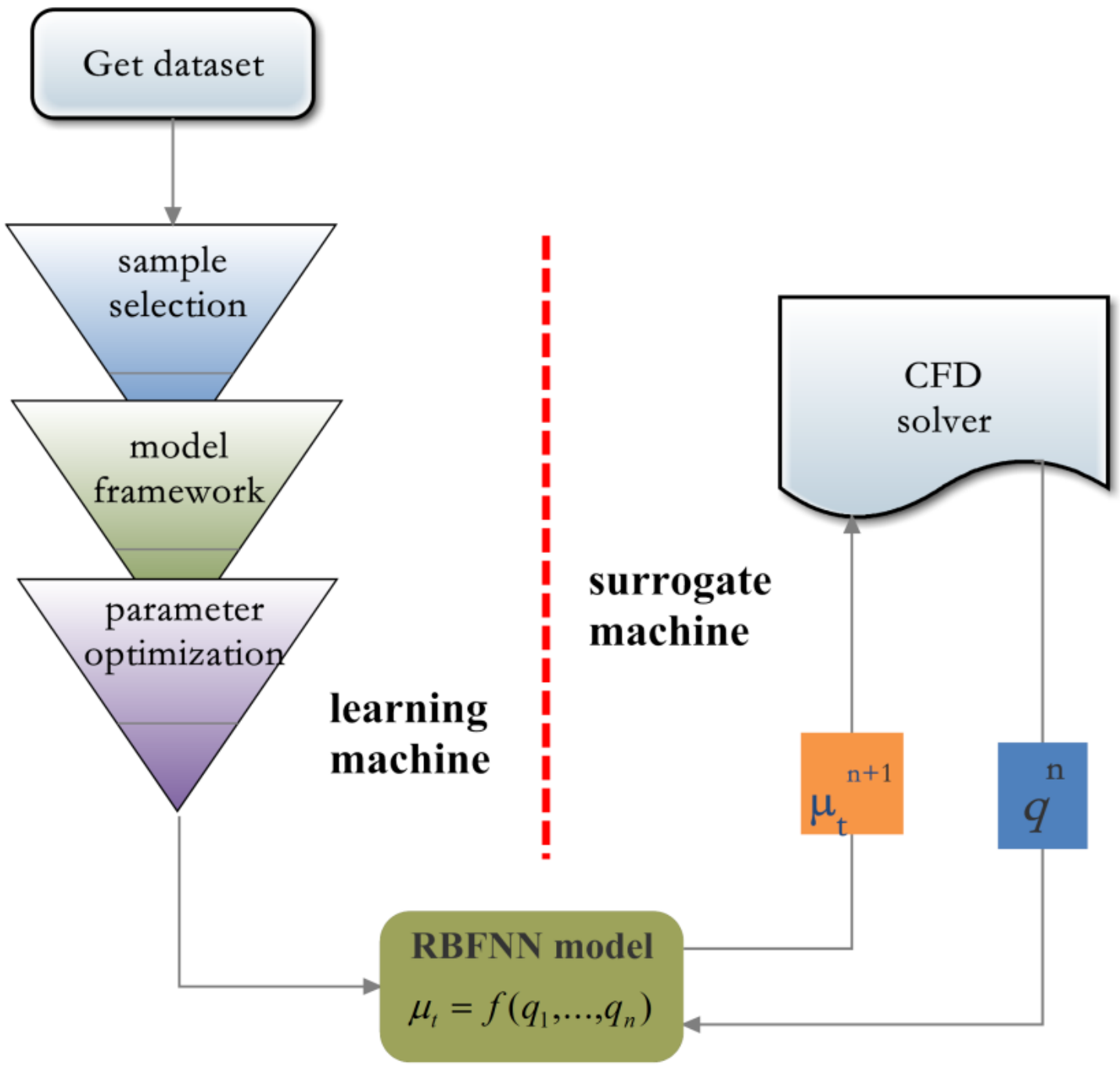

FIGURE 1. Flow chart for building the learning machine and surrogate machine where ***q*** means input feature.

## 2.2. Modeling strategy

The selection of modeling strategies depends on the specific problem. A typical strategy is the local model based on the grid topology, which approximates the output by neighboring nodes. This strategy can only be applied to fixed grid topology and needs large numbers of local models caused by overfull nodes. Another strategy computes the output according to the freestream condition like Mach number and angle of attack, etc. This strategy constructs a mapping to modal coefficients by extracting some vital modes. In essence, this projection-based model is a database between the freestream condition and output, which is hard to generalize to different geometry for lack of information of flow field. *Different from the above two strategies, the model construction process in this paper combines freestream condition and local mean variables as well as some location information together. In this way, compared with the local model, the amount of proposed model is cut down dramatically but the dimension is increased to some degree. We tried to build a model with acceptable dimension and desirable generalization ability.*

High Reynolds number means thin boundary layer. Thus, *the gradient and data range of eddy viscosity is very large along the normal direction of wall*. As a consequence, direct modeling may lead to many outliers. To avoid this problem, we referred to the idea of Xiao [15] and Gamahara [24] and divide the whole flow field into different zones according to the normal distance from the wall. Furthermore, although the partition modeling method takes effect to a certain extent, it does not reflect the truth that the small errors of eddy viscosity in near wall region may lead to large discrepancies in skin friction. In order to highlight the fitting weight of near wall region, an exponential function $F_s = e^{\sqrt{d/d_{\min}}} - 2$ was introduced in the modeling process, where $d_{\min}$ is the minimum value of normal distance from the wall in a specific zone. Another method we tried but failed to handle the large data range is using logarithmic transformation before modeling and then making inverse transformation. But this will also make the error between model output and truth value magnified exponentially. Therefore, outliers are still unavoidable and can be easier to emerge while the high accuracy is hard to guarantee for the whole data range. Another advantage of partition is the feature and model parameters can be tailored for different zones. For example, the number of hidden neurons can be increased relatively in boundary layer, and the entropy can be selected as one feature for wake region [27].

## 2.3. The artificial neural networks

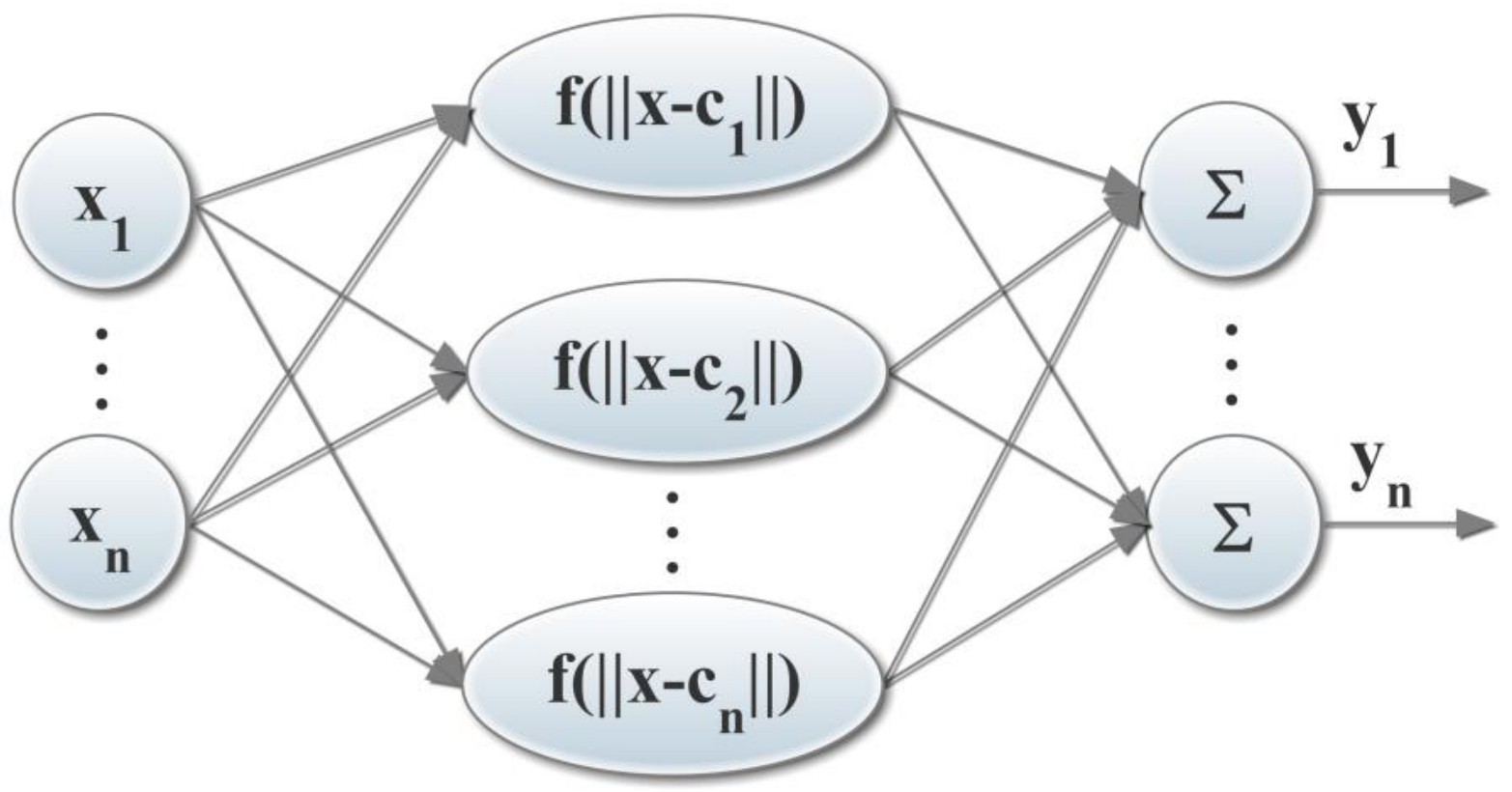

FIGURE 2. Framework of one hidden layer neural networks with double inputs and outputs.

Machine learning can be used to build a model from dataset by some algorithms, which has the ability of judgment and prediction. As one typical algorithm of machine learning, neural networks are built according to mutual connection among brain neurons, which have been widely applied to pronunciation and image recognition.

Generally, compared with the compact model, more hidden layers and neuron units is beneficial to improve modeling results, but the risk of overfitting and low generalization is also increasing. Considering the balance of accuracy and generalization, one hidden layer neural networks is adopted in this paper. The radial basis function was proved to be a great approximator [28] and has been applied to the PDEs solution [29], flow field reconstruction [30] and model of nonlinear unsteady aerodynamics [31-32]. A typical radial basis function neural networks consisting of input layer, one hidden layer and output layer is shown in figure 2. *The input layer is formed by sample features which include freestream conditions, mean flow variables and derivatives and so on, see table 1. There are 80 neurons in hidden layer and the Gaussian basis function is adopted.* The effect of number of neurons was studied in our research. The results indicate that the fitting accuracy improved slightly from 80 to 100, while the accuracy of 60 neurons is not high enough.The output layer is the eddy viscosity of each sample. To get compact datasets, the whole data points were scaled linearly to [-1 1] before training procedure.

TABLE 1. The flow features used as regression input where $\boldsymbol{S}$ the strain rate, $\boldsymbol{\Omega}$ the rotate rate, $\alpha$ the angle of attack, $\|\cdot\|$ the matrix norm. For convenience, the entropy was redefined as $S' = p/\rho^{\gamma} - 1$ in this paper.

| feature | description | sign |
|---|---|---|
| 1 | horizontal velocity | $u$ |
| 2 | density | $\rho$ |
| 3 | normal wall distance | $d$ |
| 4 | dimensional analysis | $d^2\boldsymbol{\Omega}$ |
| 5 | exponential function | $F_s$ |
| 6 | projection of free stream to normal direction of streamline | $\mathrm{sgn}(y)[-v + u\tan(\alpha)]$ |
| 7 | velocity direction | $\arctan[\lvert v/u \rvert]$ |
| 8 | entropy | $S'$ |
| 9 | strain rate | $\|\boldsymbol{S}\|/(\|\boldsymbol{S}\|+\|\boldsymbol{\Omega}\|)$ |

For a dense grid, the adjacent cells might have very similar flow information. Thus, if each cell is taken as one sample, there will be many redundant samples for similar flow cases, which can increase the training time. In addition, the distribution of sample space is imbalanced due to different grid densities in the whole domain.

Consequently, the model performance is inclined to those denser regions. Facing these two problems, this paper performed sample selection to decrease unnecessary samples, aiming at approximating the original sample space by less but more representative samples. Specifically, the algorithm is shown as follows:

1. Make sure the expected sample number $K$ of sample group $S$ and constant $\lambda$ ($\lambda > 1$) as well as $\delta$.Then choose a data point from the whole dataset $T$ randomly as the first one of sample group, $k = 1$.

2. Compute the minimum relative distance between data point $T_l$ and every sample $S_m$ ,i.e. $\min(\sum_{n=1}^{N}\left|\frac{T_{l,n} - S_{m,n}}{\max(T_{l,n}, S_{m,n})}\right|, m = 1, 2...k)$, where $N$ is the feature number. If the minimum relative distance is larger than $\delta$, then $T_l$ is chosen as a new sample, $k = k + 1$; else discarded.

3. Recycle the last step for the whole data points in $T$.

4. If $k = K$, then stop; else $\delta = \delta / \lambda$, repeat from step 2.

## 2.4. Parameter optimization

Once the sample group is determined, the parameters of center and width in each neuron can be obtained by various algorithms like gradient descent (GD), orthogonal least square and recurrent least square and so on [33-35]. The equations of gradient descent optimization algorithm are shown as follows:

$$c_j^{n+1} = c_j^{n} - \eta_c \cdot \nabla c_j \tag{2.1}$$

$$\sigma_j^{n+1} = \sigma_j^{n} - \eta_\sigma \cdot \nabla \sigma_j \tag{2.2}$$

where the gradients in above equations were calculated by

$$\nabla c_j = -2 w_j \sigma_j \sum_{i=1}^{K} e_i \boldsymbol{\Phi}(x_i)(x_i - c_j) \tag{2.3}$$

$$\nabla \sigma_j = w_j \sum_{i=1}^{K} e_i \boldsymbol{\Phi}(x_i) \left\| x_i - c_j \right\|^2 \tag{2.4}$$

The learning ratio $\eta_c$ and $\eta_\sigma$ were set to be 0.01 in this paper.

To avoid unreasonable values of model parameters [36], the centers were limited in the range of sample space during the optimization process, and the width is assigned as 0.01 if negative. The optimal weight can be obtained by GD or pseudo-inverse as following:

$$\boldsymbol{w}^{n+1} = (\boldsymbol{\Phi}^T \boldsymbol{\Phi} + \lambda(|\boldsymbol{w}|)^{\dagger})^{-1} \boldsymbol{\Phi}^T \boldsymbol{y} \tag{2.5}$$

where $|\boldsymbol{w}|$ is a diagonal matrix with diagonal element $|w_1|,|w_2|,...,|w_M|$ and $\dagger$ is the generalized inverse.

For multi-extreme value problem, GD is influenced by initial value and falls into local minimum. In order to contain more information, the clustering result is taken as the initial value. Clustering is a method used to divide the dataset into several disjoint subsets. There are various clustering methods and the distance calculating formula is dependent on the specific problem [37-40]. For less similarity between the model centers, the *K*-means clustering in [41] was used in this paper.

The loss function is the objective function for parameter optimization, like L1 (one norm) loss function and L2 (square norm) loss function, etc. It is found that although the L2 loss function can achieve higher training accuracy, the convergence of the NS equation is difficult to be guaranteed after the model is embedded in the CFD solver. As such, the L2 loss function with L1 constraint is adopted as the final objective function,

$$L = \frac{1}{2}\sum_{i=1}^{K}(y_i - f(x_i))^2 + \lambda\sum_{j=1}^{N}|w_j| \tag{2.6}$$

where $\lambda$ is between $0.01 \sim 0.05$.

## 3. Example and analysis

### 3.1. Code validation

The validation of SA model is performed for subsonic flow over NACA0012 airfoil [42] and transonic flow over RAE2822 airfoil [43]. The specific freestream conditions are shown in table 2. The adopted mesh is generated by Pointwise 18.0, with the height of first grid $d_{y^+<1} = 4.0\times10^{-6}$.

TABLE 2. The airfoil flow conditions for validation.

| Airfoil | case | $\alpha$ (/$^o$) | Re | Ma |
|---|---|---|---|---|
| NACA0012 | case1 | 0 | $3\times10^6$ | 0.15 |
| | case2 | 10 | $3\times10^6$ | 0.15 |
| | case3 | 15 | $3\times10^6$ | 0.15 |
| RAE2822 | case9 | 2.8 | $6.5\times10^6$ | 0.73 |
| | case10 | 2.8 | $6.2\times10^6$ | 0.75 |

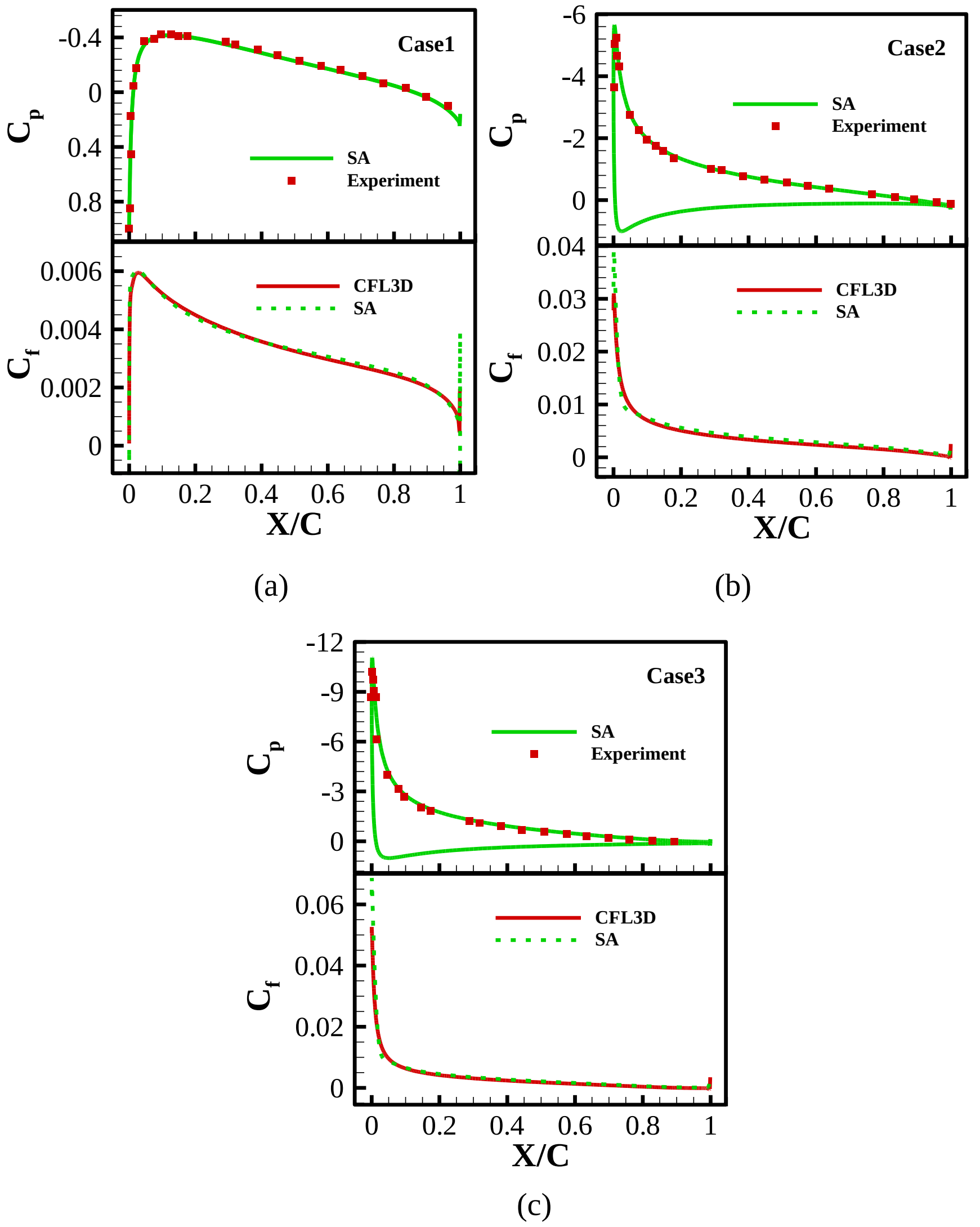

FIGURE 3. The comparison of surface pressure (upper) and skin friction coefficient (lower) in different cases of NACA0012 airfoil.

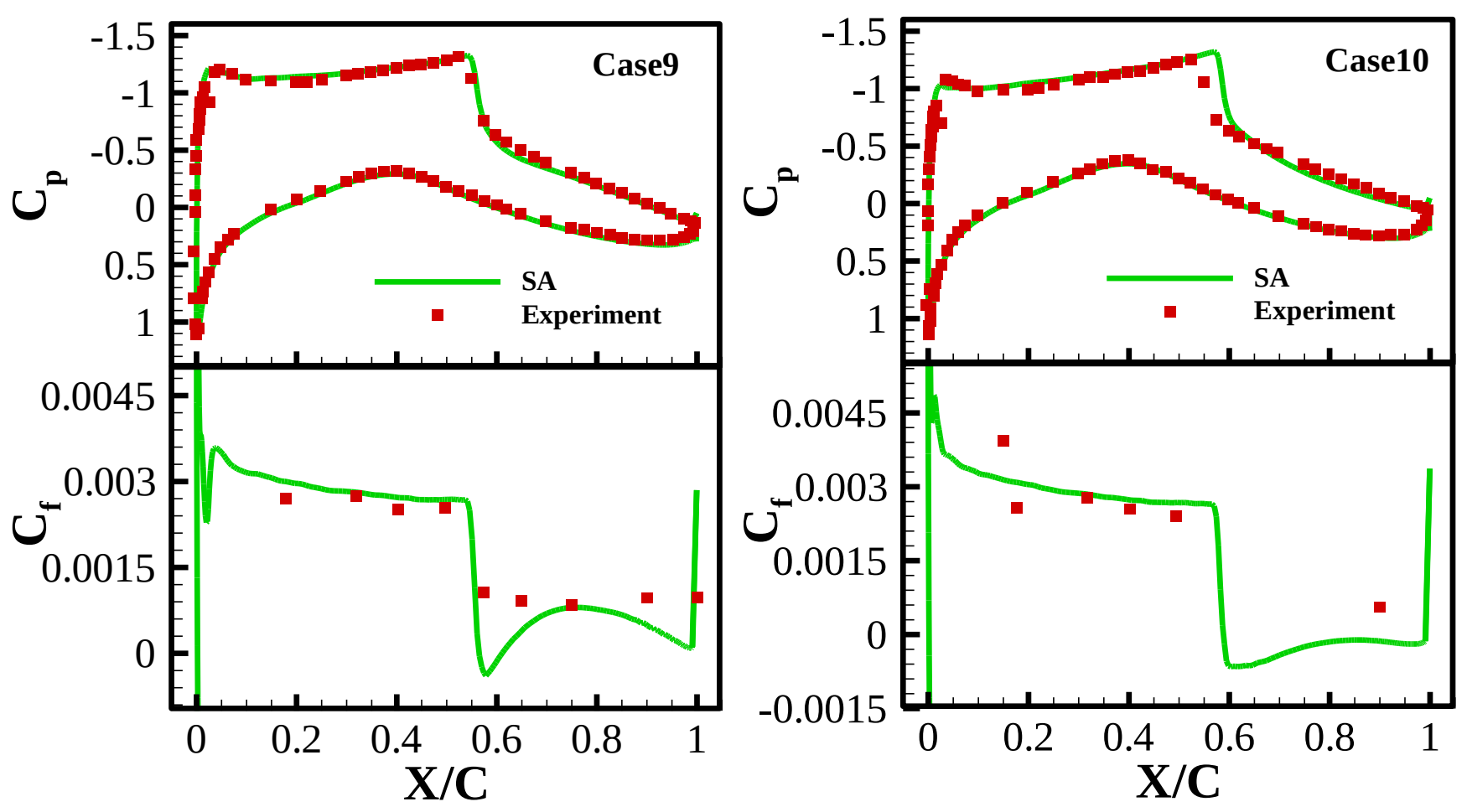

(a) (b)

FIGURE 4. The comparison of surface pressure (a) and skin friction coefficient on upper surface (b) in different cases of RAE2822 airfoil.

The pressure coefficient and skin friction coefficient around NACA0012 airfoil at three angles of attack agree well with the experimental results and CFL3D results respectively, see figure 3. Figure 4 shows the comparison of case9 and case10 with the experimental results around RAE2822 airfoil. In case 10, the shockwave location moved downstream slightly. It should be emphasized that, the skin friction coefficient in figure 4(b) is non-dimensionalized by the local boundary layer freestream condition rather than the incoming flow condition [44].

## 3.2. Result

Like [8], the results calculated by CFD solver with SA model [45] were regarded as the true value in this paper. Subsonic steady flows over NACA0012 airfoil, NACA0014 airfoil and RAE2822 airfoil were investigated at fixed Reynolds number $\mathrm{Re}=3\times10^6$. The results shown below mainly includes skin friction coefficient, velocity and eddy viscosity profiles along the normal direction of wall and eddy viscosity contour. Specifically, six monitoring locations are selected from both upper and lower surface of airfoil, which are $X/C=0.09894$, $X/C=0.4887$, $X/C=0.8213$ on the upper surface and $X/C=0.09894$, $X/C=0.5001$, $X/C=0.8006$ on the lower surface, see figure 5.

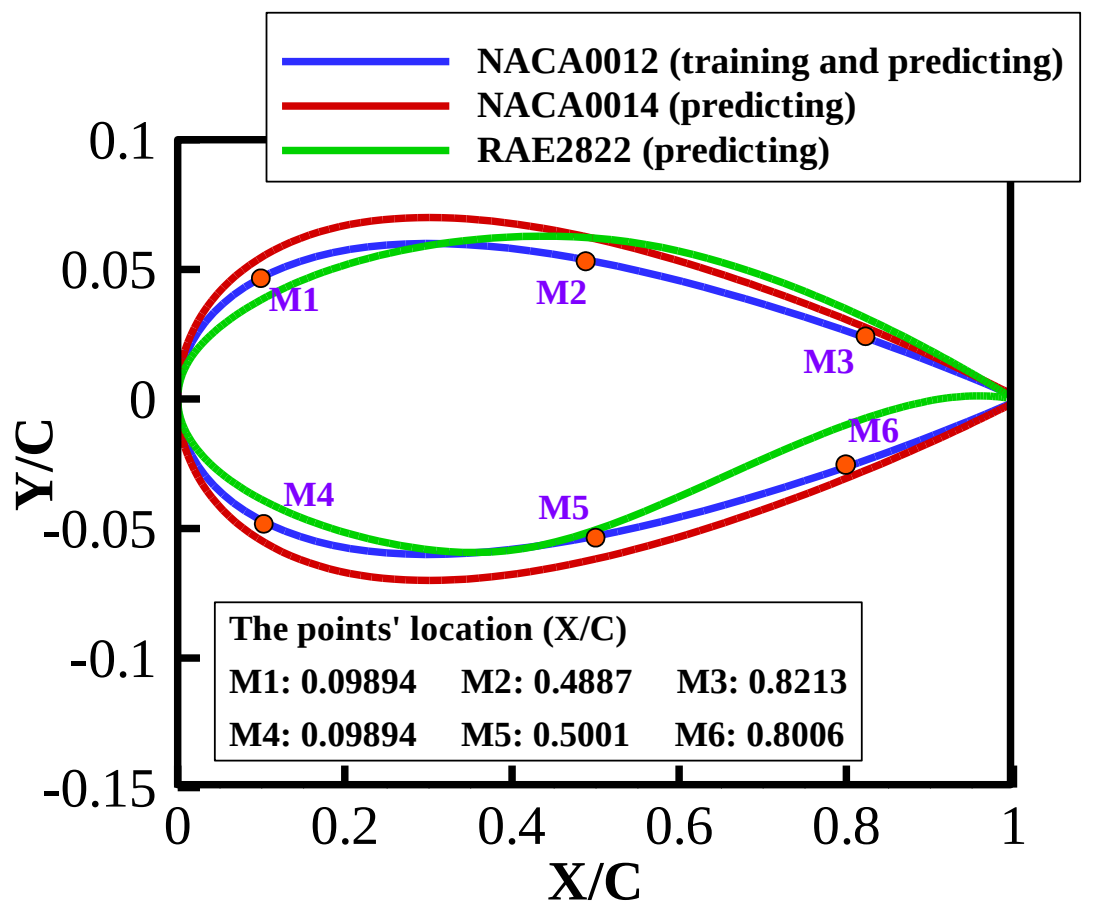

FIGURE 5. The adopted airfoils for training and predicting the RBFNN model.

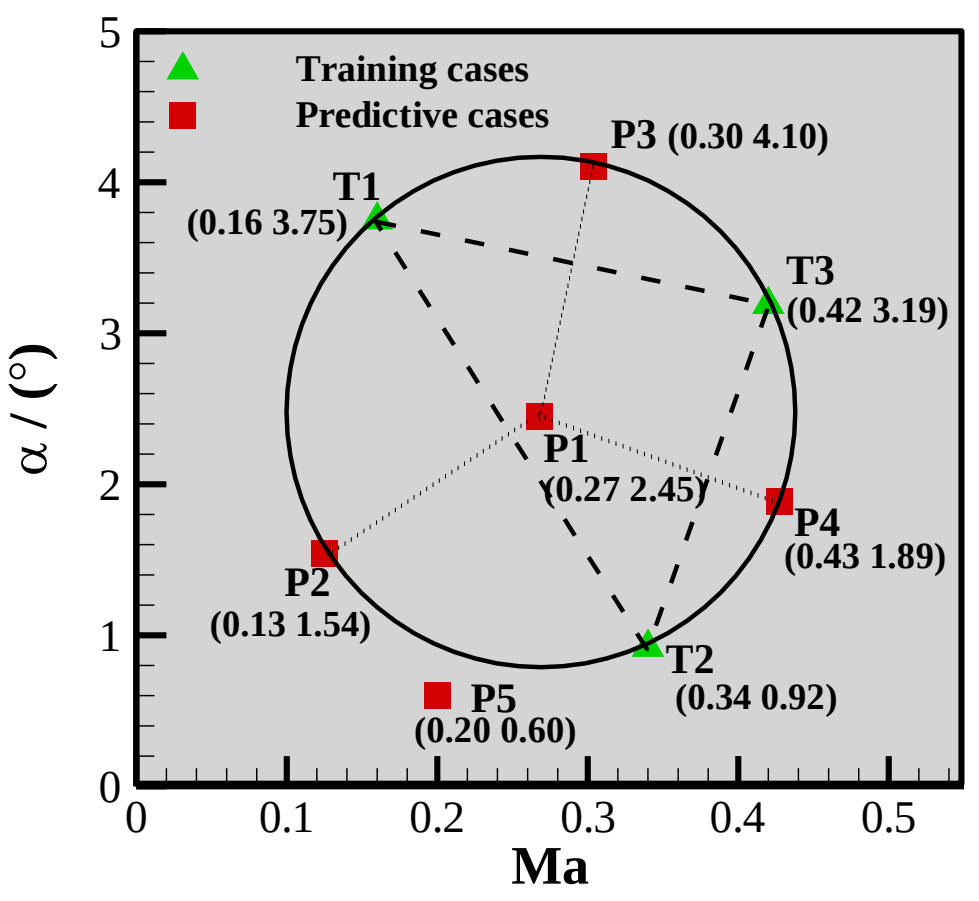

FIGURE 6. The adopted training and predictive cases.

*Three flow fields over NACA0012 airfoil is chosen as training cases (T1, T2, T3),* which was sampled through Latin hypercube sampling method (LHS) with Mach number (Ma) from 0.1 to 0.5 and angle of attack ($\alpha$) from 0 to 5 degrees. *The predicting cases include both interpolation/extrapolation of flow states of NACA0012 airfoil and typical states of two different airfoils, by which the generalization of proposed model will be demonstrated.* The predicting cases were chosen as the circumcenter P1 (interpolation) and the intersections of circumcircle with three perpendicular bisectors P2-P4 (extrapolation) as well as P5 (extrapolation) outside the circumcircle, see figure 6.

The hybrid grid was adopted, of which the first layer height in boundary layer is $d_{y^+<1} = 8.0\times10^{-6}$ and the increasing rate is 1.2. Taking NACA0012 airfoil as an example, the whole domain and the local grid near the leading edge is shown in figure 7(a) and 7(b), respectively.

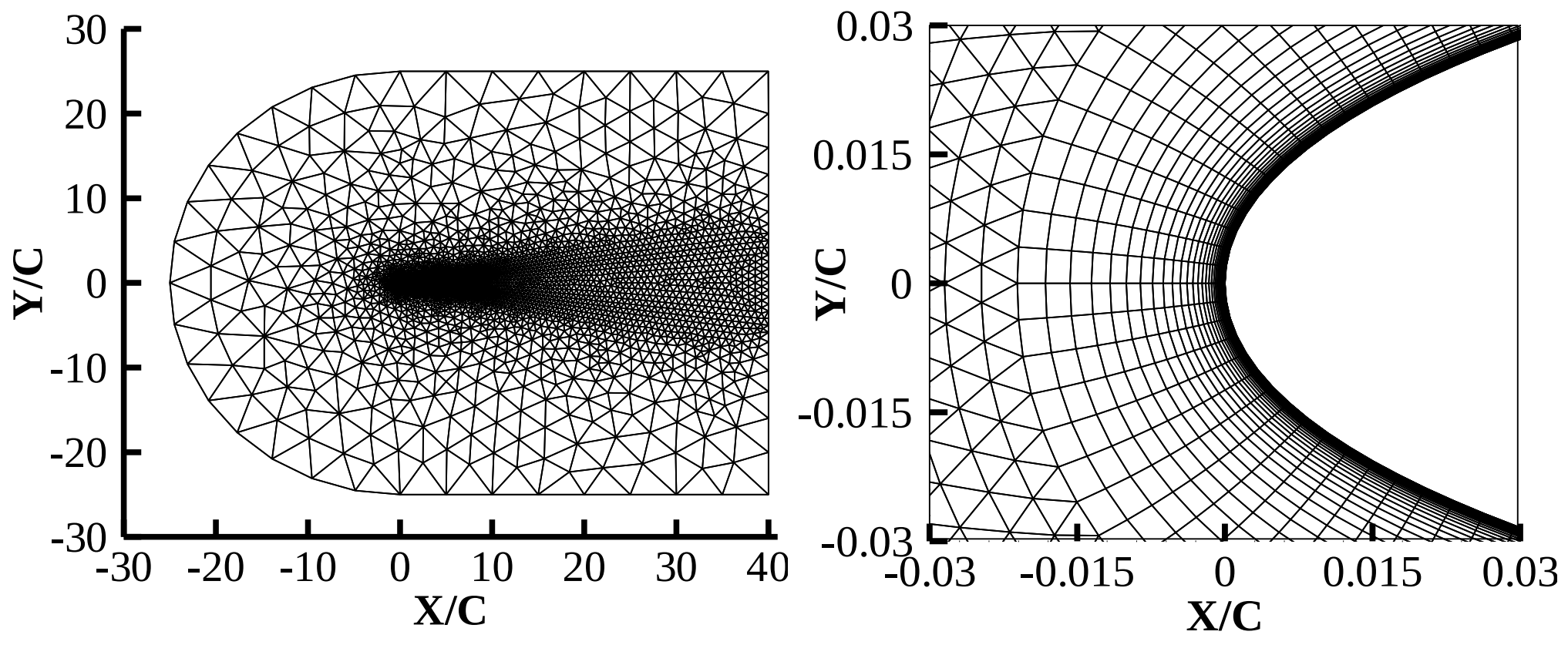

(a) (b)

FIGURE 7. The adopted mesh for NACA0012 airfoil.

### 3.2.1. The training cases

The skin friction coefficient of three training cases is almost identical with the true value, with slight error on lower surface and rear part of upper surface, see figure 8. Taking T2 case as the example, the eddy viscosity contour agrees well with the true value except for wake region, see figure 9. As for the eddy viscosity profile at monitoring locations (figure 10), obvious errors can be observed from the peak value down to near zero. In this region, poor agreement is mainly caused by low sensibility of input to output. Specifically, the change of mean flow variables is flat while the change of eddy viscosity is still sharp along the normal direction of wall. But these large discrepancies have little impact on the velocity profile, see figure 11.

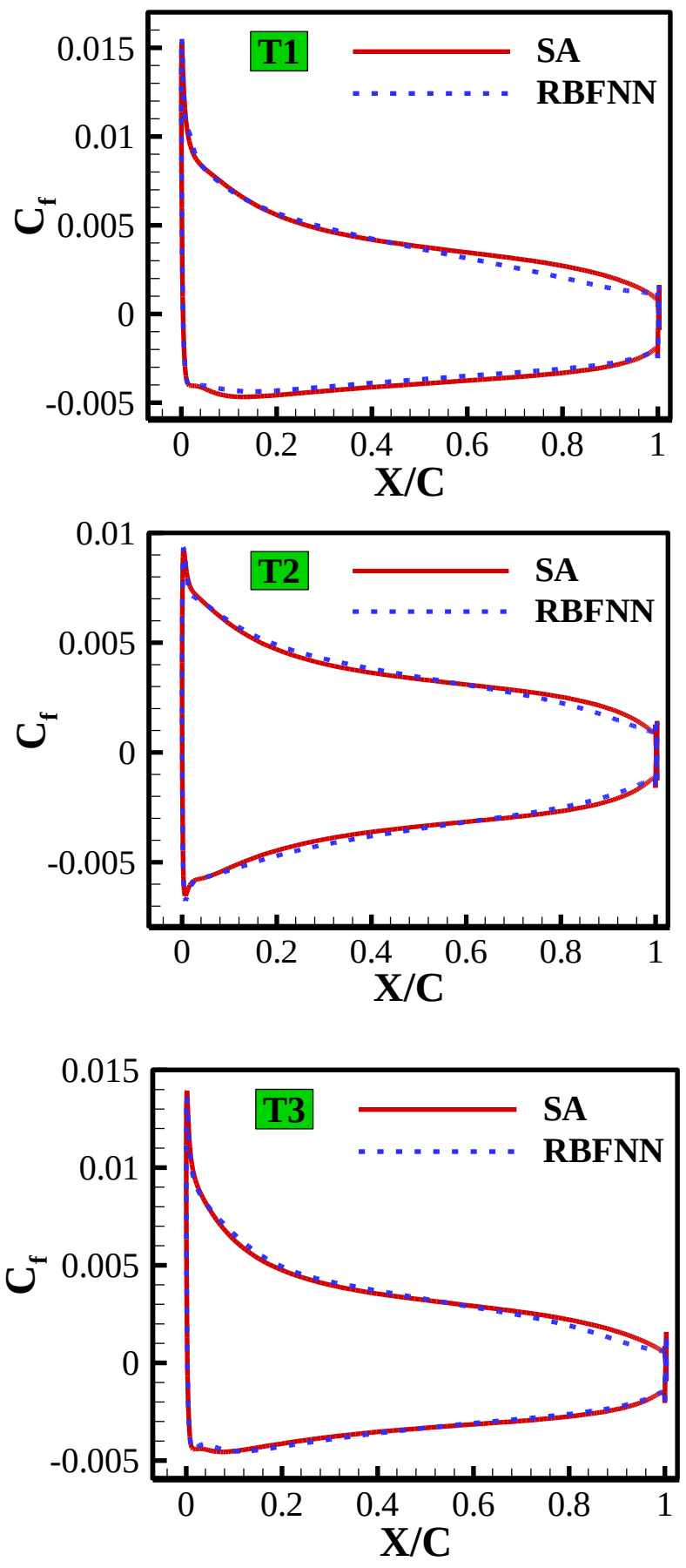

FIGURE 8. Comparison of the skin friction coefficient.

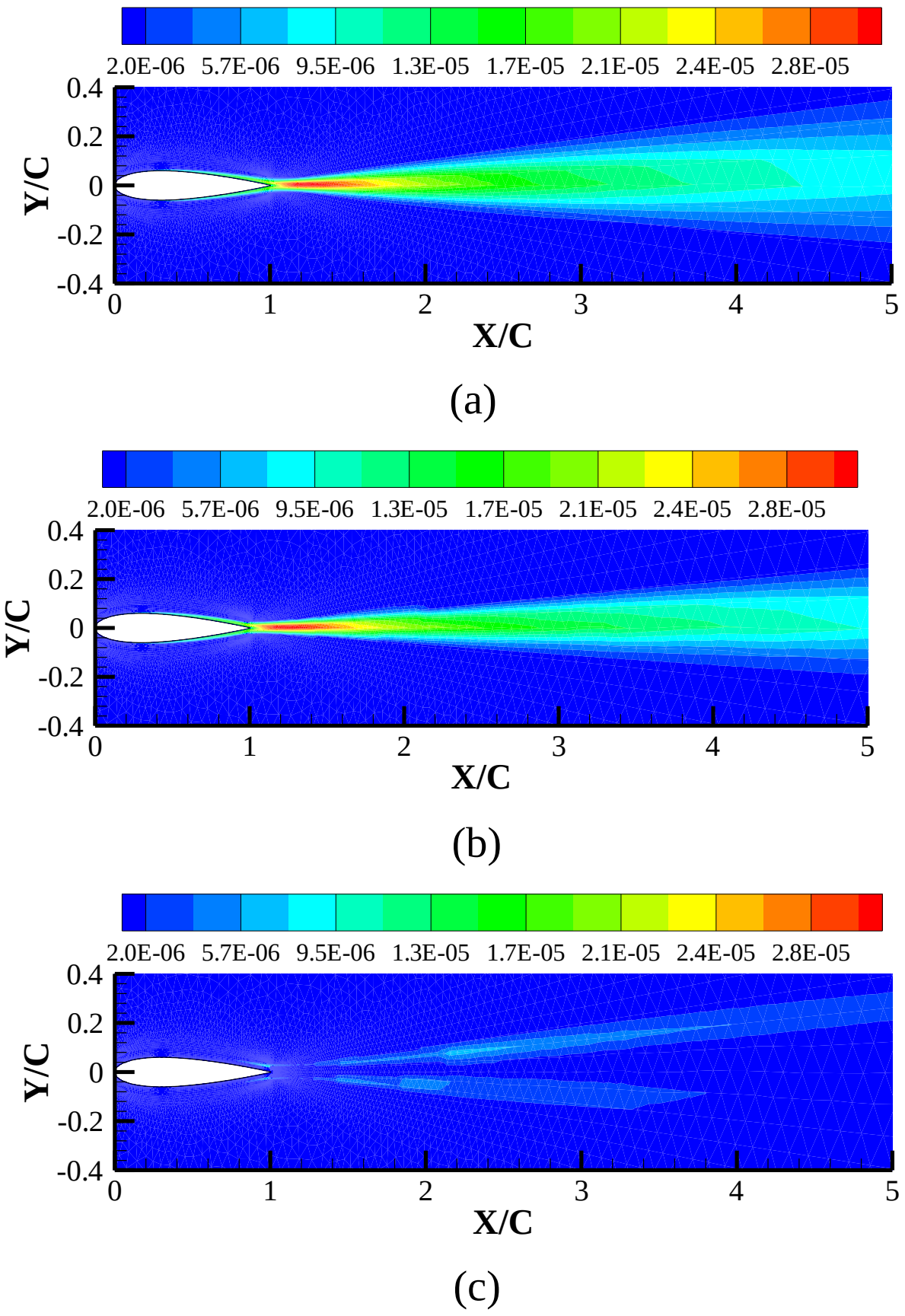

FIGURE 9. The contour of eddy viscosity at T2 case calculated by (a) SA model (b) RBFNN model and (c) the error contour.

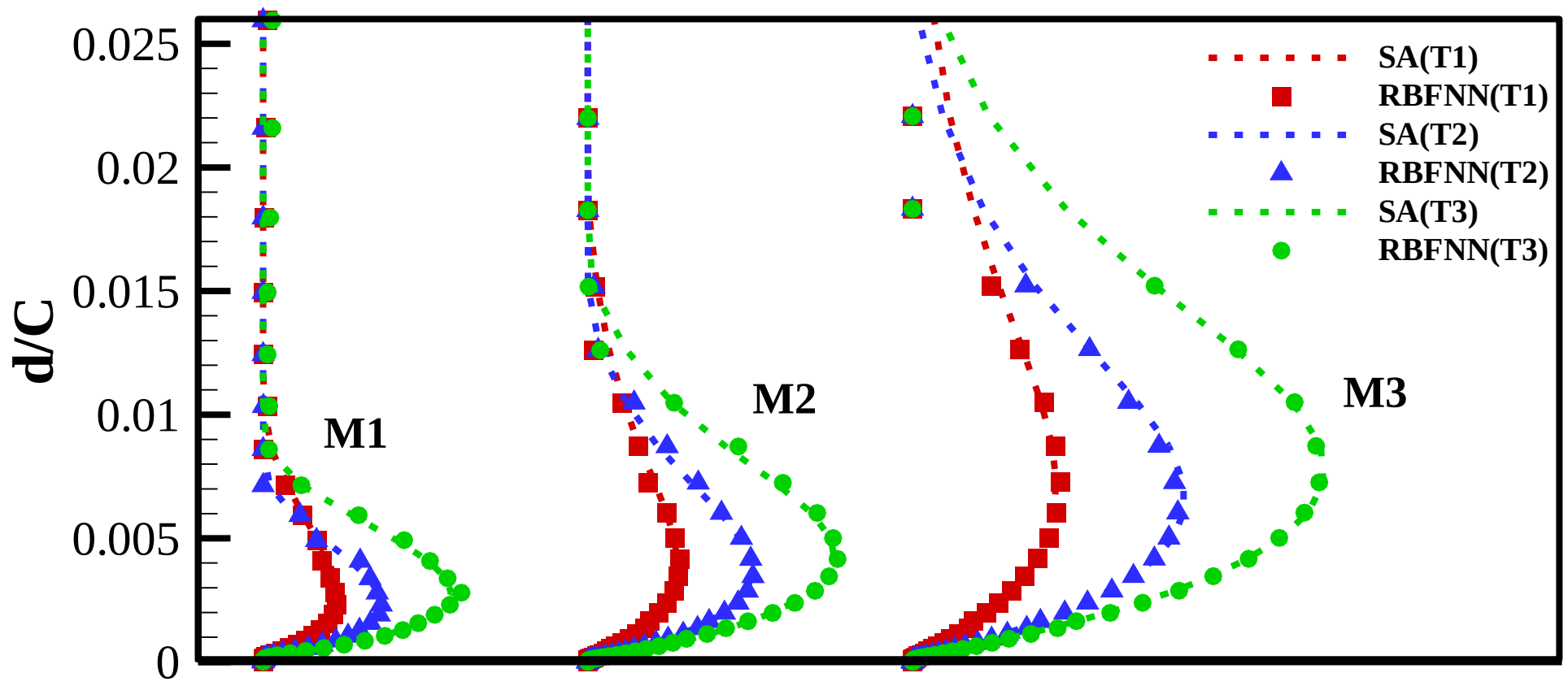

(a)

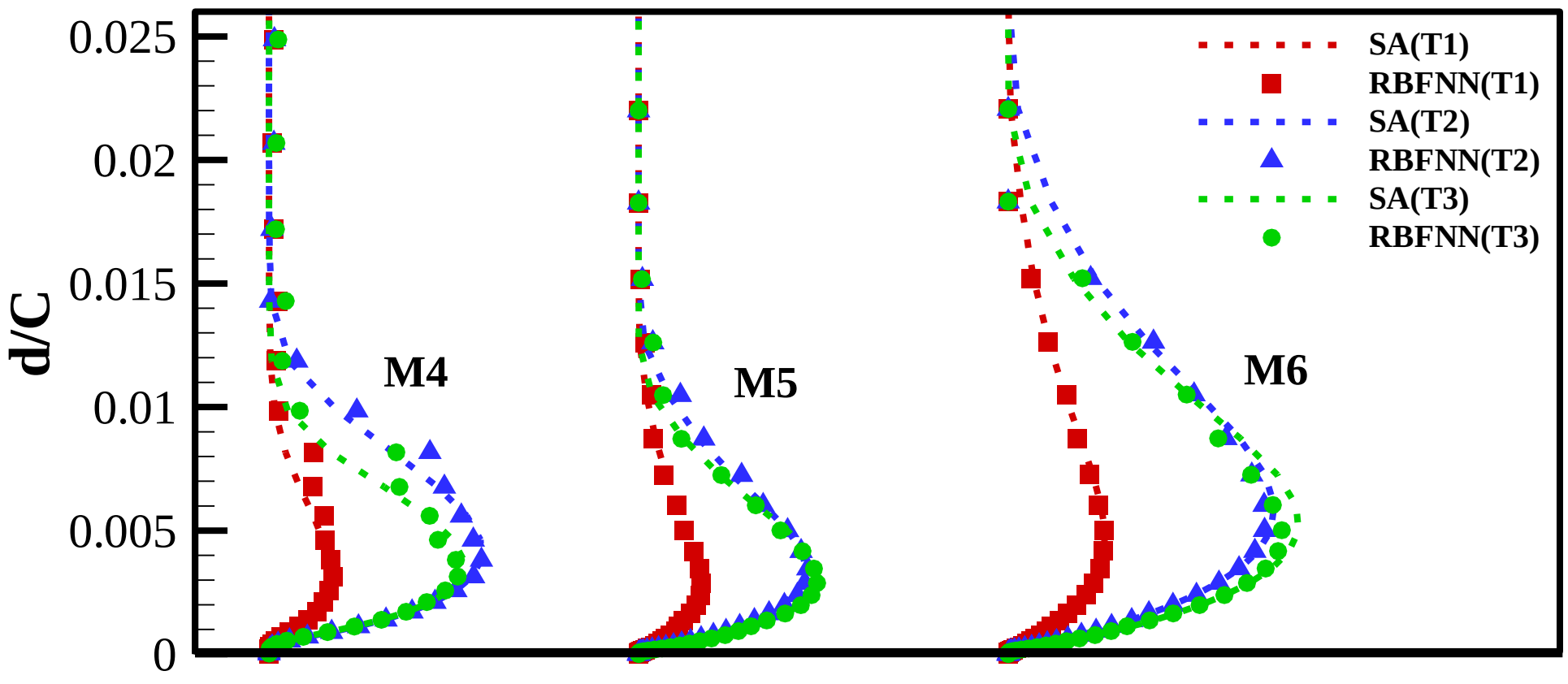

(b)

FIGURE 10. The eddy viscosity profile of training cases at monitoring points along the normal direction of wall (a) upper surface (b) lower surface. For clearance, the profiles of upper surface and lower surface at M1and M4 are magnified by three and six times, respectively.

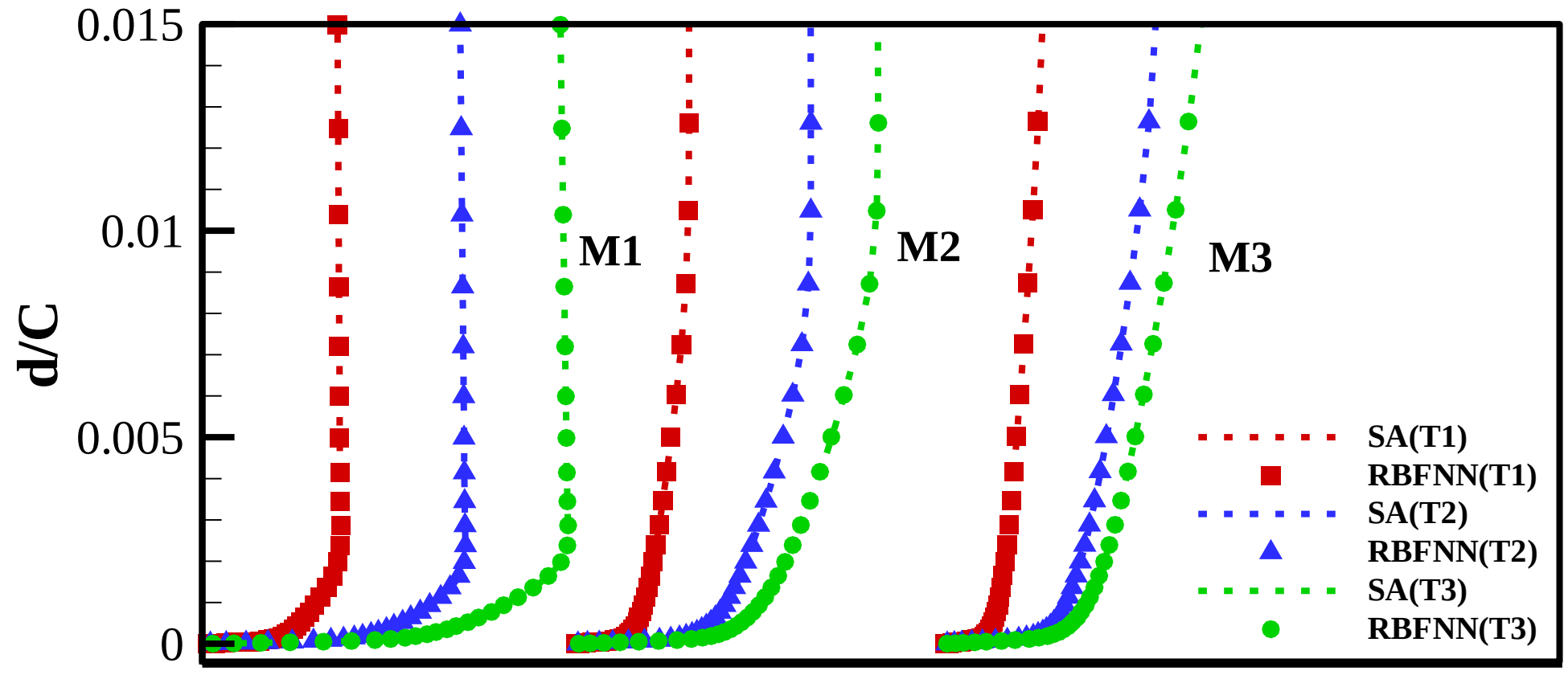

(a)

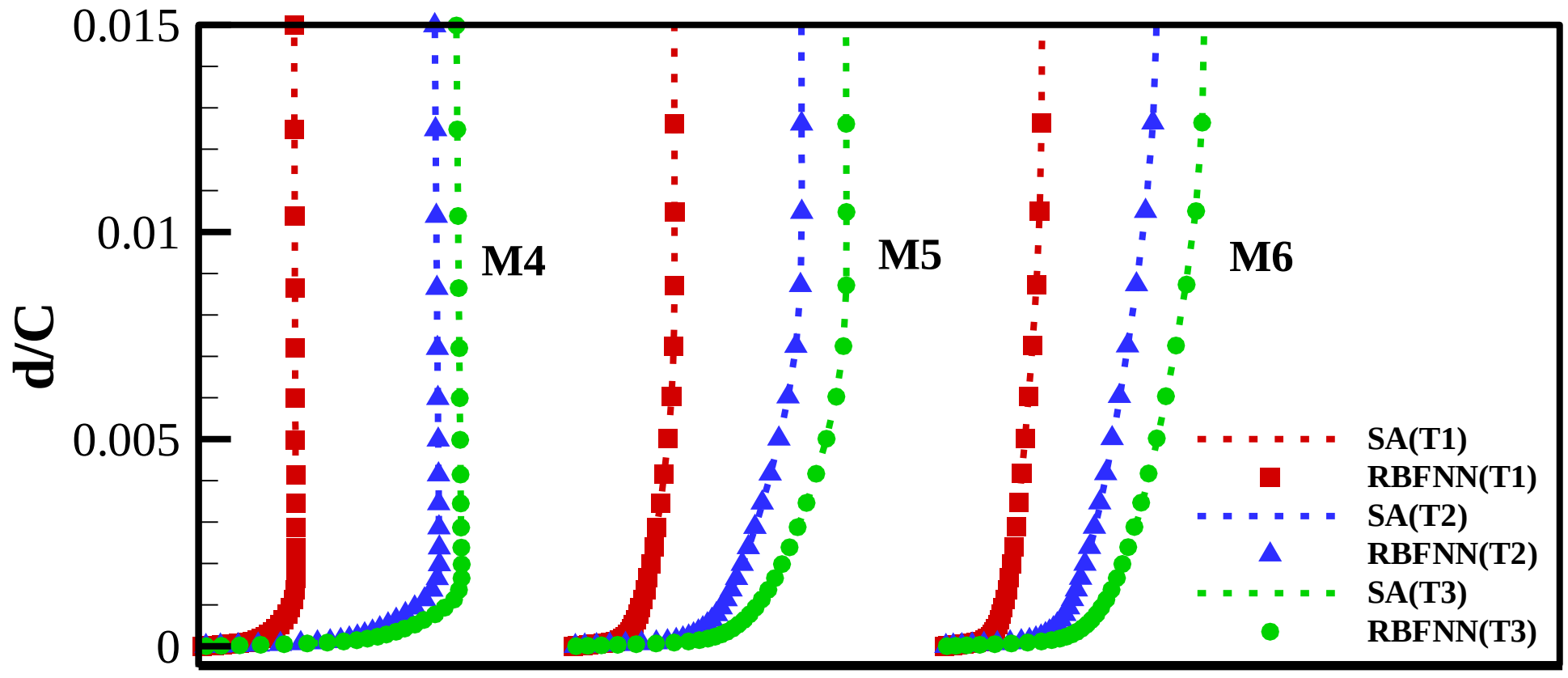

(b)

FIGURE 11. The velocity magnitude profile of training cases at monitoring points along the normal direction of wall (a) upper surface (b) lower surface.

### 3.2.2. The predicting cases

**Part Ⅰ NACA0012 airfoil**

The skin friction coefficient of five predicting cases also shows good agreement except P2 case which has the largest deviation from training cases, see figure 12. The contour of P4 case and eddy viscosity profiles have similar agreement as training cases, see figure 13-14. Excellent agreement of velocity profiles along the normal direction of wall at monitoring locations is shown in figure 15. Both lift coefficient and drag coefficient is shown in figure 16, with agreement of lift coefficient almost identical. The mean relative error of drag coefficient is only 1.79% and the maximum corresponding to P2 case is 0.0007, basically caused by the skin friction error.

***Although training dataset consists of only three cases, while each case contains abundant local flow information, which ensures the diversity of samples.*** Besides, for every divided zone, the proposed model is a global model, which approximates the output by the information of whole zone rather than neighboring nodes like local model. And, there is no qualitative difference between training and predicting cases, when the flow field has neither shock waves nor separations. Therefore, *the proposed model shows good agreement when generalized to both interpolated and extrapolated cases.*

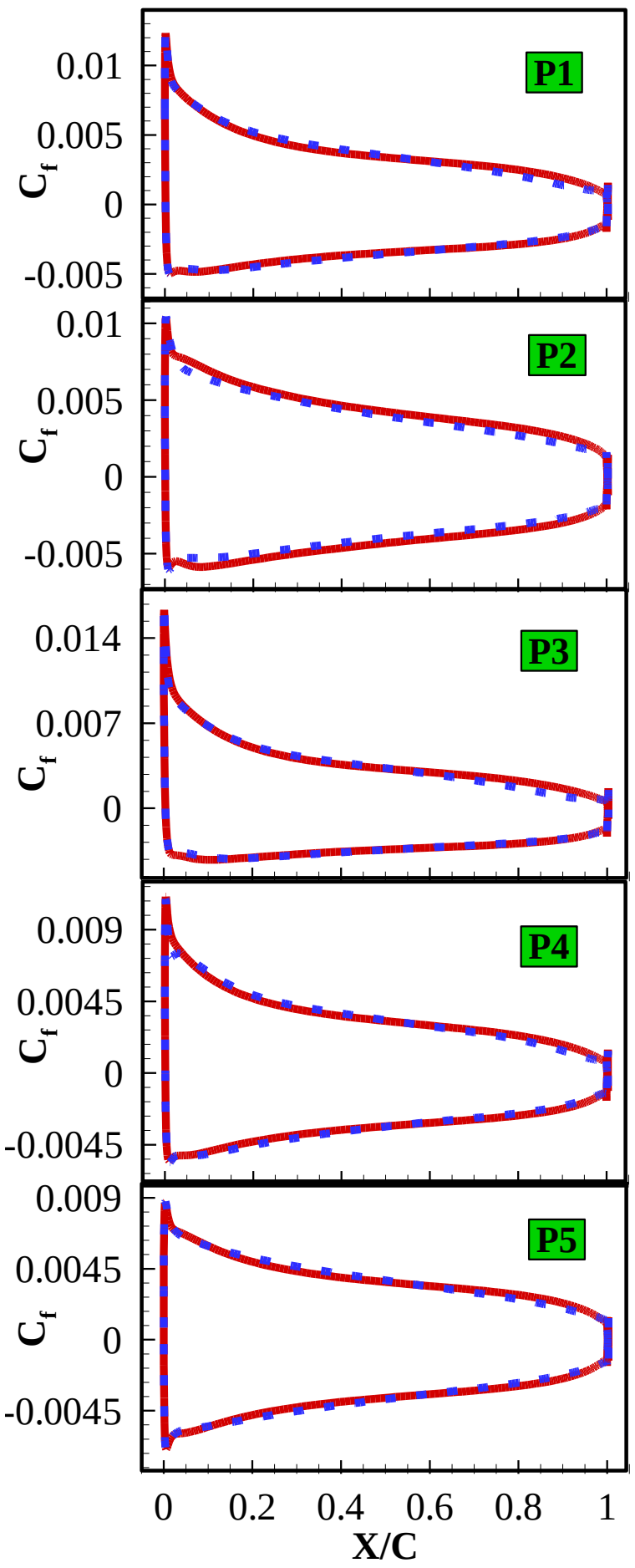

FIGURE 12. Comparison of SA (solid red line) and RBFNN (dashed blue line) for predicting cases. Not to scale.

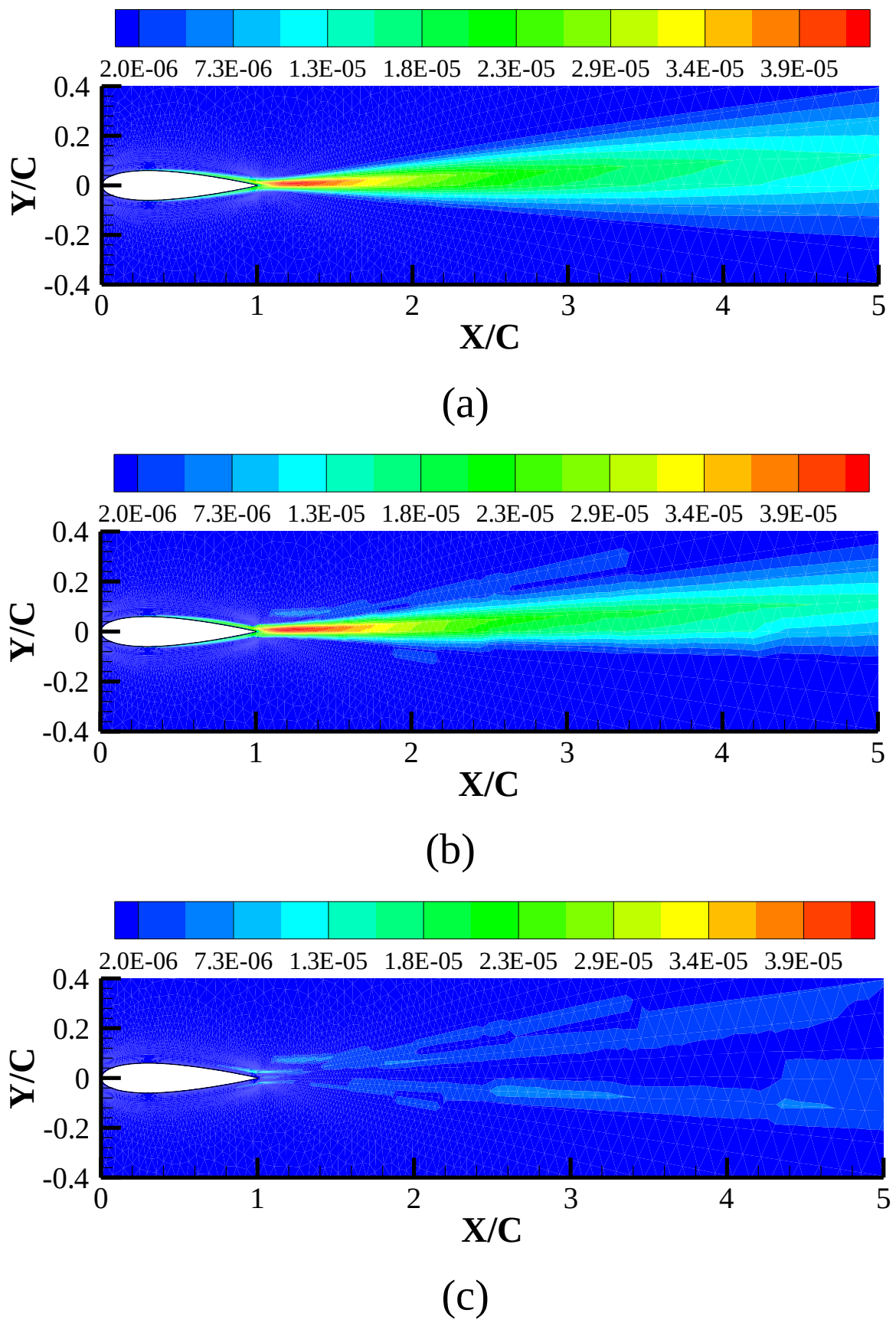

FIGURE 13. The contour of eddy viscosity of P4 calculated by (a) SA model (b) RBFNN model and (c) the error contour.

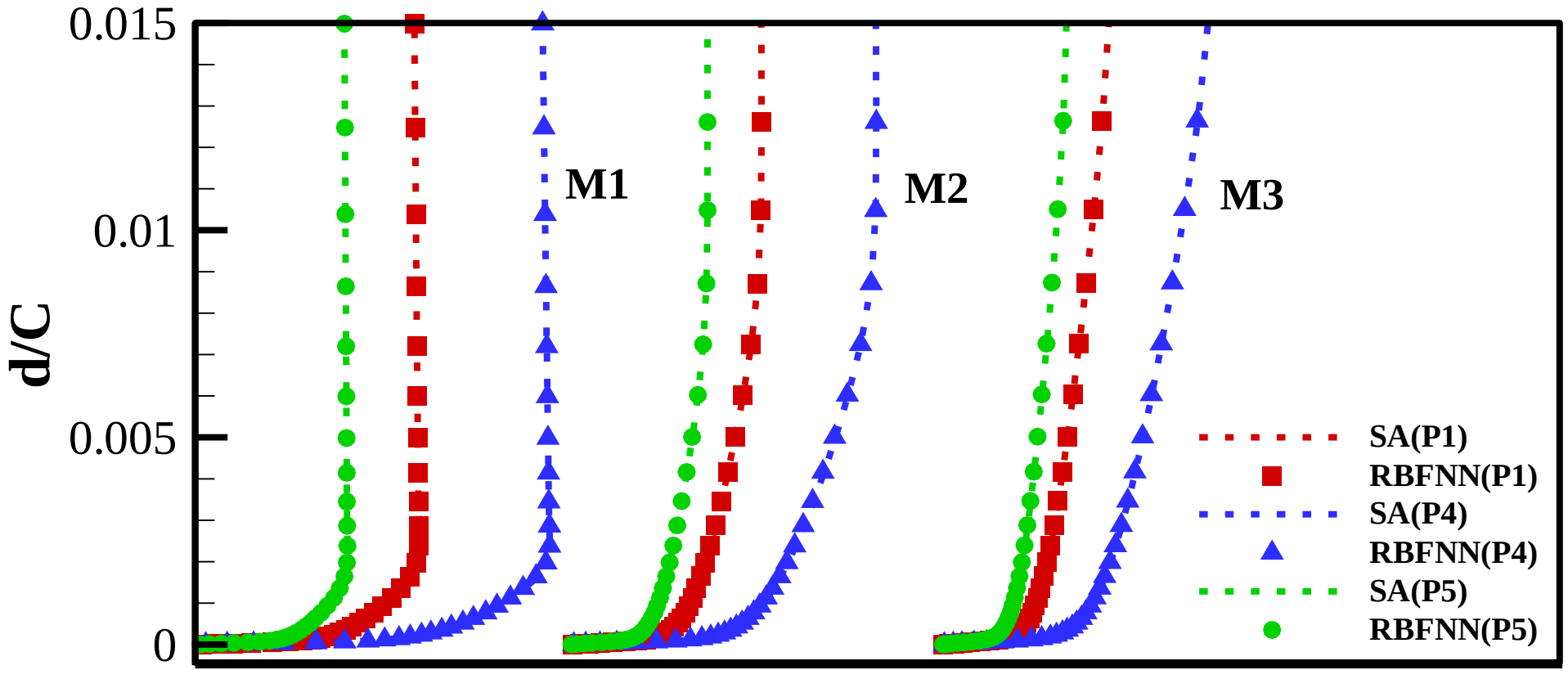

(a)

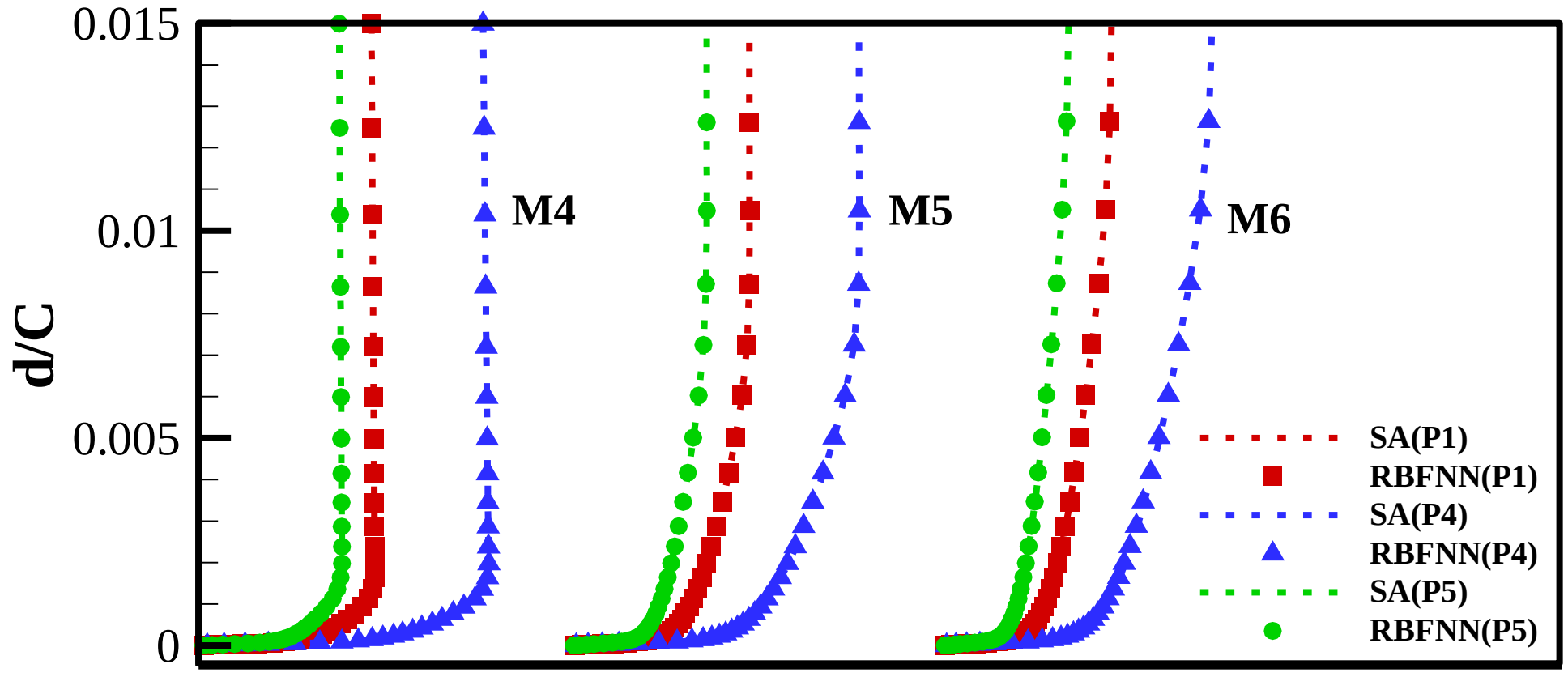

(b)

FIGURE 14. The velocity magnitude profile of three predicting cases at monitoring points along the normal direction of wall (a) upper surface (b) lower surface.

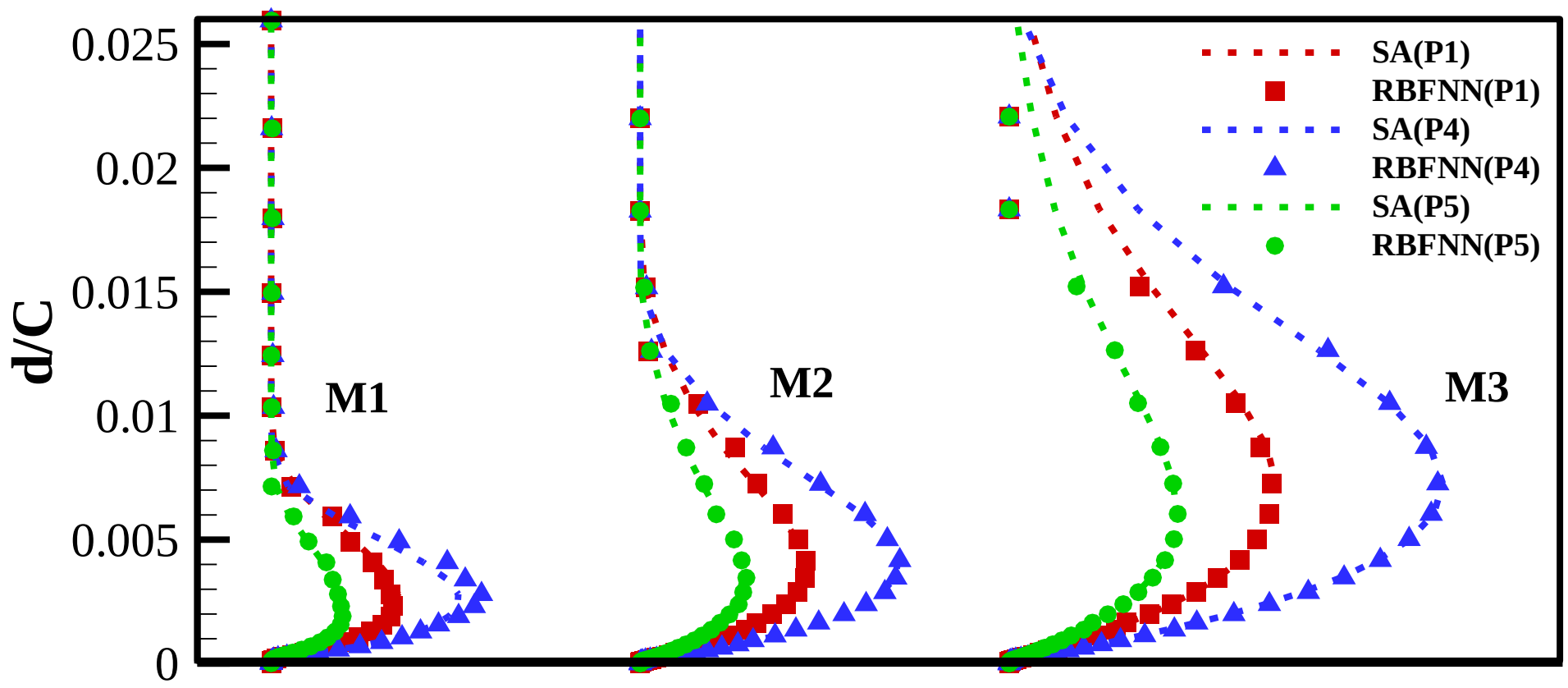

(a)

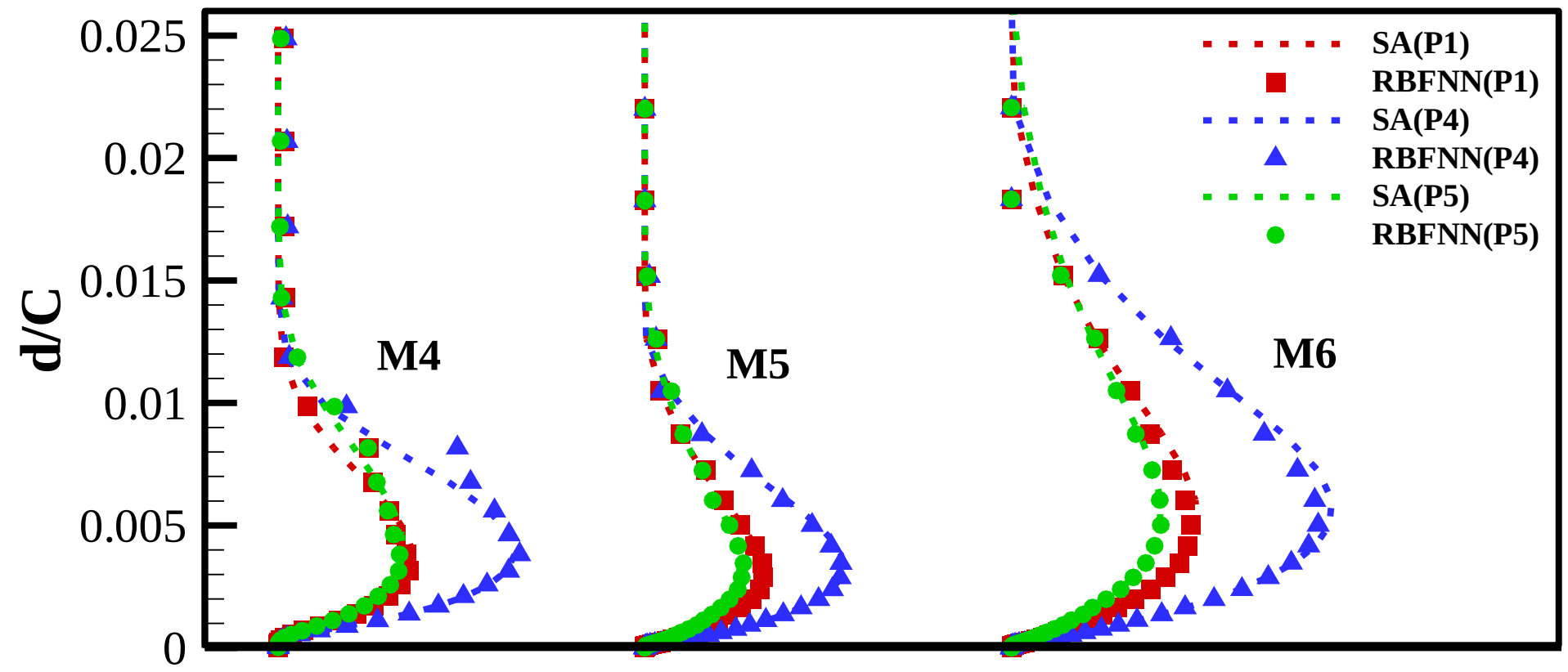

(b)

FIGURE 15. The eddy viscosity profile of three predicting cases at monitoring points along the normal direction of wall (a) upper surface (b) lower surface. For clearance, the profiles of upper surface and lower surface at M1and M4 are magnified by three and six times, respectively.

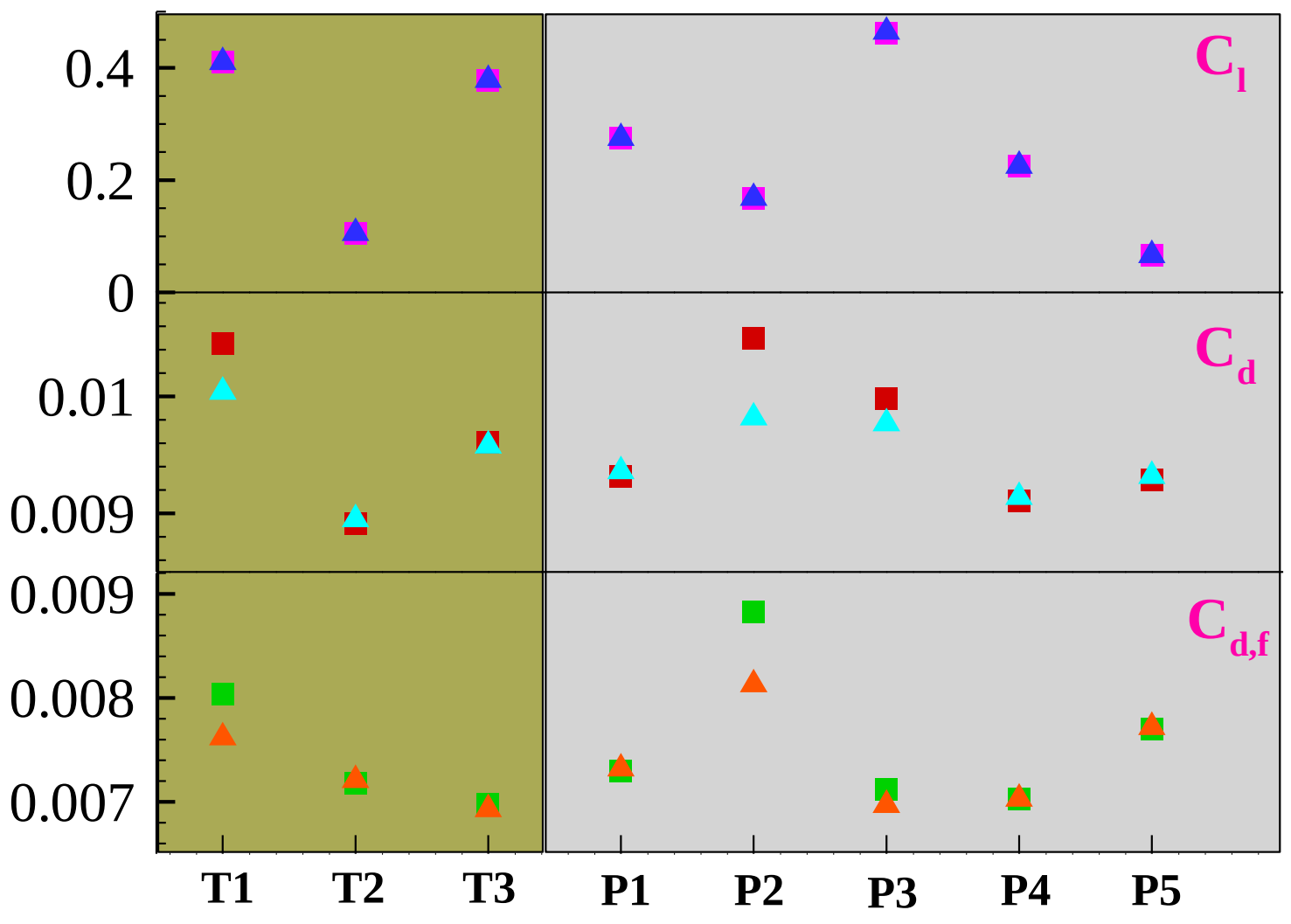

FIGURE 16. Comparison of SA (square) and RBFNN (delta) for both training and predicting cases.

**Part Ⅱ NACA0014 airfoil and RAE2822 airfoil**

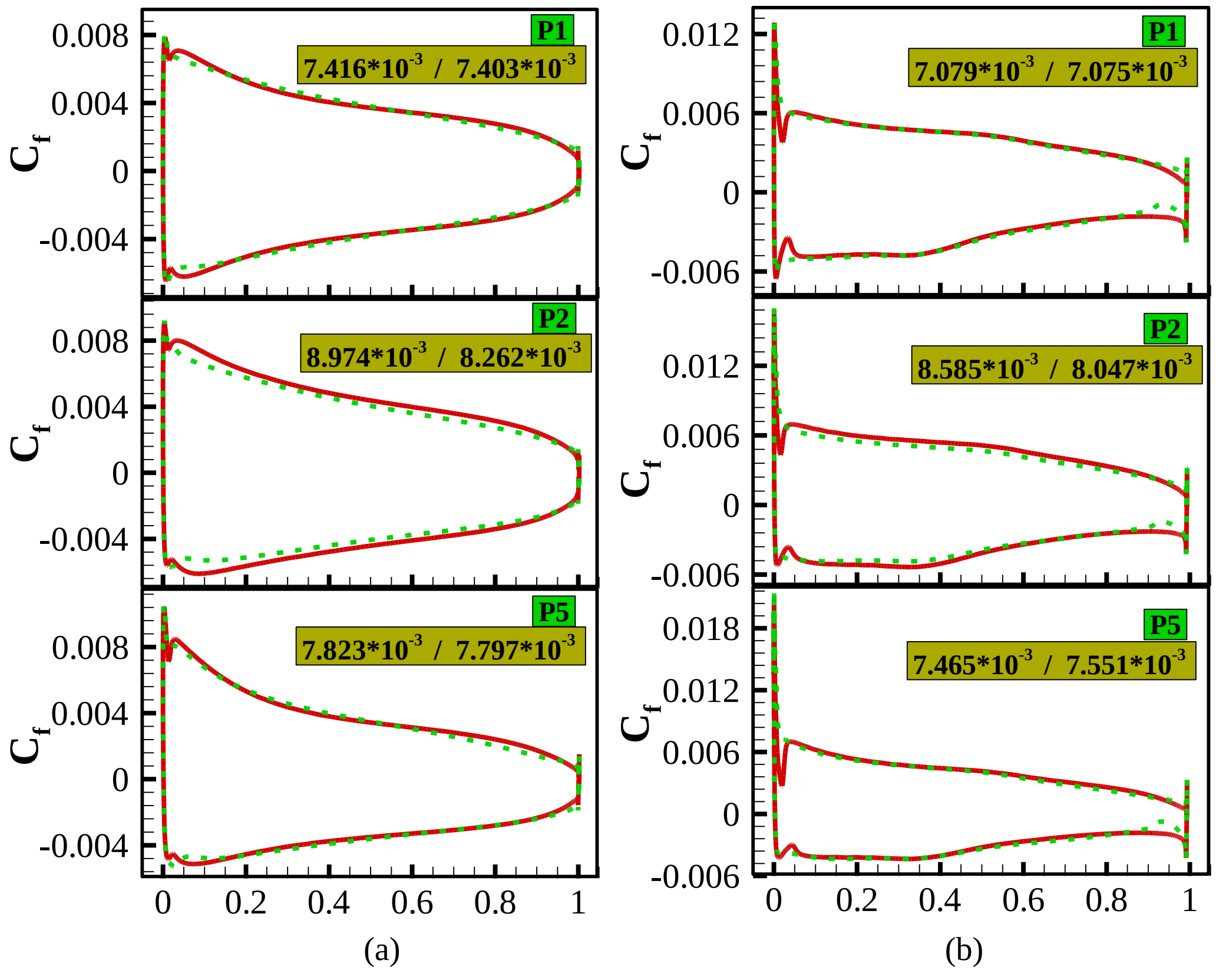

FIGURE 17. Predictions for NACA0014 (a) and RAE2822 (b) airfoil at P1, P2 and P5 cases. Not to scale. The data inside are $C_{d,f}$ values calculated by SA/RBFNN model.

*In this part, NACA0014 airfoil and RAE2822 airfoil were adopted to test the generalization of above model driven by NACA 0012 airfoil data for different airfoil*

*shapes*. Considering both the interpolation and extrapolation, P1，P2 and P5 were selected as the computing cases. The results show skin friction coefficients are in good agreement except P2 case, see figure 17. Although there are sharp shifts of residual during the computation process, CFD solver embedded with the present model still achieved satisfying convergence. The residual evolution of P5 case is shown in figure 18.

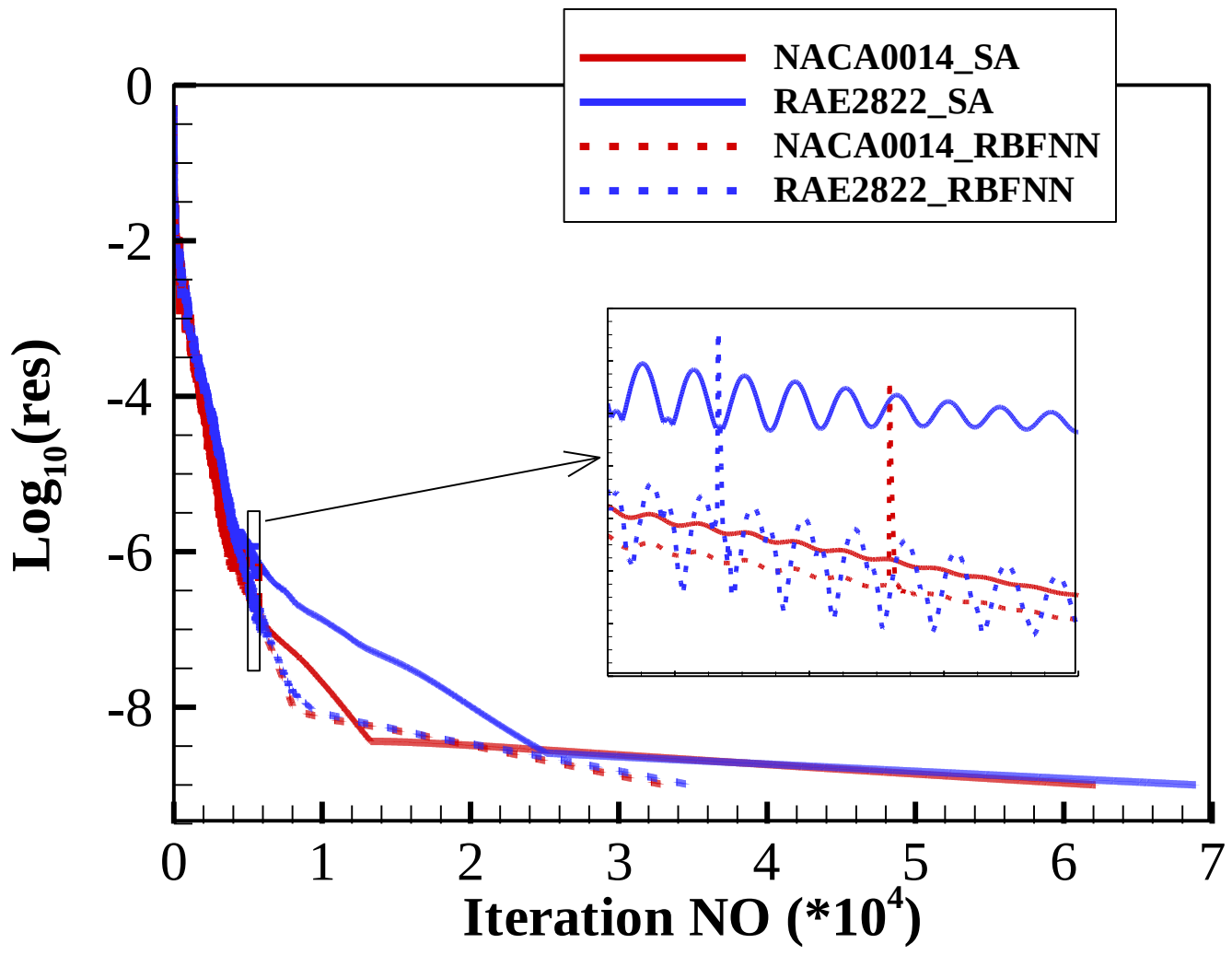

FIGURE 18. Residual evolution at P5 case.

High efficiency is also one of targets in our work. The one hidden layer neural network is a concise framework without solving transport equations. We listed the computing time of five predicting cases about NACA0012 airfoil and three predicting cases about NACA0014 airfoil and RAE2822 airfoil. For nearly all the cases, the proposed approach is more efficient, especially for those flow cases with better accuracy, see table 3.

TABLE 3. Comparison of turbulence model's computing time as the residual was down to $O(10^{-9})$. The black, green and blue data are corresponding to NACA0012, NACA0014 and RAE2822 airfoil, respectively.

| Computing time (s) | SA model | RBFNN |
|---|---|---|
| P1 | 1975.5/1135.8/1519.8 | 1111.5/680.5/867.2 |
| P2 | 858.9/430.7/984.5 | 717.82/641.1/738.6 |
| P3 | 1995.6 | 760.2 |
| P4 | 1864.3 | 714.4 |
| P5 | 1708.1/971.6/1349.8 | 693.50/693.0/787.7 |

## 4. Conclusions and future work

In this paper, based on three training cases of turbulent flows over NACA0012 airfoil, the radial basis function neural network was adopted to model the eddy viscosity for subsonic attached flows. By comparing the proposed approach with original SA model, the accuracy and generalization capability to different airfoils and flow states are validated. The conclusions are stated as follows:

(1) By partition and building the model separately, the outliers caused by large data range can be decreased effectively, which is good to obtain satisfying accuracy in vital domains. And coupled with Navier-Stokes equations, the proposed approach also achieves the final convergence.

(2) The present model is a kind of global model with appropriate dimensions, which achieves high accuracy and generalization while only needs a few training cases. For both the training cases and predicting cases, the velocity profile and skin friction distribution agree well with the SA model, which demonstrates the promising prospect of machine learning methods in future works about turbulence modeling.

(3) The proposed approach is more efficient than original SA model. On the one hand, the one-hidden layer neural networks with about a hundred neurons are a concise framework without complex calculation. On the other hand, less iteration steps are needed for achieving the final convergence standard.

This paper is still a preliminary work toward modeling high Reynolds number turbulent flows with data-driven methods. Separated flows and other more complex turbulent flows will be further investigated in future works.

**Acknowledgements**

This paper is mainly supported by the National Science Fund for Excellent Young Scholars (no. 11622220), 111 project of China (B17037) and National Natural Science Foundation of China (11572252).